\documentclass{epl}

\usepackage{graphicx}
\usepackage{amsmath}
\usepackage{amsfonts}

\renewcommand{\r}{\mathbf{r}}
\newcommand{\eff}{{\text{eff}}}

\newlength{\GraphicsWidth}
\title{Charge inversion of colloids in an exactly solvable model}

\author{Gabriel T\'ellez}
\institute{Departamento de F\'{\i}sica, Universidad de Los Andes,
A.A.~4976, Bogot\'a, Colombia}


\begin{document}

\maketitle

\begin{abstract}
We study a two-dimensional model for a long cylindrical stiff
charged macroion immersed in a charge-asymmetric electrolyte with
charge ratio $+2$/$-1$. The model is integrable and it allows an exact
analytical determination of the effective charge of the macroion,
which characterizes the electrostatic potential at large distances
(compared to the screening length) from the macroion. At at high
coulombic coupling, this model predicts charge inversion: for a highly
negatively charged macroion, the effective charge could become
positive, indicating an overscreening of the macroion by the divalent
counterions. By studying the behavior of the coions and counterions
density profiles close to the macroion, we show that the counterion
condensation threshold is shifted to a lower value in absolute
value. This plays an important role in the charge inversion
phenomenon.
\end{abstract}



The determination of the effective interactions between charged
macroions immersed in electrolyte solutions is a central topic in
colloidal science~\cite{Verwey-Overbeek}. Based on the linear
Debye--H\"uckel (DH) theory, for a very long and thin (zero radius)
cylindrical macroion, with linear charge density $e/\ell$, the reduced
effective potential at a distance $r$ is $y(r)= 2 \lambda K_0(\kappa
r)$, where $\lambda=l_B/\ell$, $K_0$ is the modified Bessel function
of order 0, $y=e \psi/(k_B T)$ with $\psi$ the electric potential, $e$
the elementary charge, $T$ the temperature and $k_B$ is Boltzmann
constant. Here, $\lambda$ is the reduced linear charge density,
expressed in units of the inverse of the Bjerrum length $l_B=e^2/(k_B
T\varepsilon)$, where $\varepsilon$ is the electric permittivity of
the solvent. The Debye screening length is $\kappa^{-1}=(4\pi l_B
\sum_{\alpha} n_{\alpha}^b v_{\alpha}^2)^{-1/2}$, where $n_{\alpha}^b$
is the bulk density of the microions of the electrolyte of type
$\alpha$ and $v_{\alpha}$ their valence. For highly charged macroions,
the linear approach is inappropriate. A first improvement over the
linear theory is to use the mean field nonlinear Poisson--Boltzmann
(PB) equation. Under this approximation, the effective potential, at
large distances, is again a screened potential: $y(r) \sim 2
\lambda_{\eff} K_0(\kappa r)$ for $\kappa r \gg 1$, but the prefactor
$\lambda_{\eff}$ is not anymore the bare charge $\lambda$ of the
colloid, but it is known as the effective or renormalized
charge~\cite{Alexander,Trizac-Bocquet-Aubouy-PRL}.

PB approach is adequate~\cite{Levin-rep,Tellez-Trizac-sat} if the
coulombic couplings between microions are small, for example for a
two-component electrolyte: $v_1^2\Gamma \ll 1$, $v_2^2\Gamma \ll 1$,
$|v_1 v_2|\Gamma \ll 1$ where $\Gamma=2l_B/a$, with $a$ the average
distance between microions in the electrolyte. For large coupling, new
phenomena can occur that cannot be explained by the mean field PB
approach. The most striking one is the phenomenon of charge inversion:
a large fraction of counterions can condense into the macroion and the
resulting dressed macroion can have a charge of opposite sign than its
bare charge. Closely related to this phenomenon is the possibility of
attraction between two like-charged macroions. Several experimental
observations and studies of these phenomena have been
reported~\cite{Bungenberg, Kabanov, Butler-PRL-multivalence} and some
theoretical explanations have been put
forward~\cite{Barrat-Joanny,Ha-Liu, Perel-Shklovskii-PhysicaA,
Ulander-Kjellander, Levesque-Leger, Messina-Holm-Kremer}, for a review
see~\cite{Grosberg-colloquium}.

In this work, we report exact results for the renormalized charge of a
cylindrical macroion, valid beyond the mean field approximation, in a
whole range where the coulombic coupling can be large: $0\leq \Gamma <
2/3$. The model under consideration is an infinitely long charged
cylindrical colloid, with radius zero, immersed in a charge-asymmetric
electrolyte solution composed of microions with valences $v_1=2$ and
$v_2=-1$.  Due to the translational symmetry along the direction of
the colloid, we consider a two-dimensional (2D) model: we will be
interested only in the microions density profiles and the electric
potential along the radial direction from the colloid. Our model
predicts the possibility of charge inversion: the effective charge and
bare charge of the colloid have opposite signs when
$\lambda<\Gamma-1<0$ and $\Gamma>1/2$.

Being two-dimensional, our model describes correctly a cylindrical
macroion immersed in an electrolyte formed by cylindrical parallel
coions and counterions, with linear charge densities $v_{1,2} e/a$. As
above, the coulombic coupling is $\Gamma=2l_B/a$. Many like-rod
colloids and polyelectrolytes can be described by the 2D model
presented here. For example, synthetic polyelectrolytes such as
poly(p-phenylene) and poly(styrenesulfonate) backbones used in water
management have tunable linear charge density and a very large
persistent length (20 nm) compared to other length scales involved in
the problem. Of particular interest for biological and biophysics
process, DNA (single and double stranded) and actin filaments are
anionic cylindrical polymers that have also a large persistent length,
and as such could be, in a first approach, described by the present
model. With these cylindrical polyelectrolytes the situation described
by this model could be reproduced experimentally. Experimental
situations similar (although not exactly equal) to the one presented
here are described in~\cite{Kabanov,DeRouchey-Netz-Radle-polyDNA}. The
experiments described in~\cite{Kabanov} and the possibility of charge
inversion in those situations have important applications in gene
therapy. Besides the possible applications to polyelectrolytes, it is
well-known~\cite{KT} that the 2D Coulomb gas is also a prototype model
that can describe the physics of many interesting systems such as
vortices in a superfluid, the XY model, and dislocations in a 2D
crystal.

For a cylindrical macroion with point-like coions and
counterions, one important ingredient in the theoretical
explanations~\cite{Ha-Liu} for charge inversion is the formation of a
2D liquid of counterions in the vicinity of the surface of the
macroion and the strong longitudinal correlations between them. Our
model, being two-dimensional, does not take into account these
longitudinal correlations, but nevertheless shows that charge
inversion is possible, thus showing that the radial correlations are
enough to drive the charge inversion phenomenon.

Our 2D model with zero radius charges is stable against the collapse
of pairs of opposite charges provided that $0<\Gamma<1$. For technical
reasons explained below our results are valid for $0<\Gamma<2/3$. To
ensure the stability against collapse of counterions into the
macroion, we require $-1/2<\lambda<1$. The study of the effective
charge beyond those limits requires to consider a macroion with
nonzero radius and it is beyond the scope of this Letter, see
however~\cite{Trizac-Tellez-Manning, Samaj-guest-charges}. Provided
$-1/2<\lambda<1$, the introduction of a small hard core radius $R$
($\kappa R\ll 1$) for the macroion is an irrelevant
perturbation. These limits for $\lambda$ correspond to the Manning
thresholds for counterion condensation~\cite{Manning} derived within a
mean field approach.  In all the theoretical approaches to charge
inversion~\cite{Barrat-Joanny,Ha-Liu, Perel-Shklovskii-PhysicaA} the
counterion condensation phenomenon is crucial. This is not an
exception here. We will show that, due to the strong correlations
between microions, the threshold for counterion condensation, in the
case $\lambda<0$, is changed to $(\Gamma-1)/2$: it take place before
than predicted by the traditional Manning theory.

Exacts results for our model can be obtained due to the recent
advances in the theory of two-dimensional Coulomb systems: the exact
bulk thermodynamics of the charge-symmetric~\cite{Samaj-Travenec-TCP}
and the ($+2$/$-1$) charge-asymmetric~\cite{Samaj-asym}
two-dimensional two-component plasma are known. This has been possible
due to their relationship with the integrable sine-Gordon (sG) and
complex Bullough-Dodd (cBD) models. Using the field theoretical tools
from the sG model, exact results for the effective charge and other
quantities for a cylindrical macroion in a charge-symmetric
electrolyte have been obtained~\cite{Samaj-guest-charges}. Here, we
use the known expressions for the form-factors of exponential fields
of the cBD model to obtain results for the charge-asymmetric case.


First, let us consider the electrolyte in the absence of any
macroion. Carrying out a Hubbard-Stratonovich transformation, the
grand canonical partition function of the electrolyte can be cast as
the partition function of the cBD model, with action~\cite{Samaj-asym}
\begin{equation}
  \label{cBD-action}
  S=\int \left[\frac{1}{16\pi}|\nabla \phi(\r)|^2-z_1 e^{ib\phi(\r)}-z_2
  e^{-ib\phi(\r)/2}\right] \,d\r
\end{equation}
where $z_{1,2}$ are the fugacities of the microions and
$b=\sqrt{\Gamma}$. To give a precise meaning to the fugacities in the
cBD model the conformal normalization should be used: $\langle
\phi(0)\phi(\r)\rangle_{z_1=z_2=0}=-\ln r$. In the following we shall
denote $\langle\cdots\rangle$ an average with the cBD
action~(\ref{cBD-action}).  We now consider that a single macroion
with charge $\lambda$ is immersed in the electrolyte at the
origin. Let $Q=\lambda/\Gamma$. The density of positive and negative
microions at a position $\r$ from the macroion
are~\cite{Samaj-guest-charges, Tellez-guest-charges,
Tellez-small-r-asympt}
\begin{equation}
  \label{n-correl}
  n_{\pm}(r)=n_{\pm}^b \frac{
    \langle e^{ib Q \phi(0)} e^{ib q_{\pm} \phi(\r)} \rangle}{
    \langle e^{ib Q \phi}\rangle 
    \langle  e^{ib q_{\pm} \phi} \rangle}
\end{equation}
with $q_{+}=1$ and $q_{-}=-1/2$ respectively and $n_{\pm}^b$ are the
bulk densities, far from the colloid. The above correlation function
can be expressed as a sum over all intermediate $N$-particle states of
the cBD model as~\cite{Smirnov-book,Samaj-asym}
\begin{multline}
  \label{correl-expans}
  \langle e^{ib Q \phi(0)} e^{ib q_{\pm} \phi(\r)}
  \rangle=
  \langle e^{ib Q \phi(0)} \rangle\langle e^{ib q_{\pm} \phi(\r)}
  \rangle+\\
  \sum_{N=1}^{\infty}
  \sum_{\epsilon_1,\ldots,\epsilon_N}\int_{\mathbb{R}^N}
    \frac{\prod_{i=1}^{N} d\theta_{i}}{(2\pi)^N N! }
    F_{Q}(\theta_1,\epsilon_1;\ldots;\theta_N,\epsilon_N)
    \overline{F_{q_{\pm}}(\theta_1,\epsilon_1;\ldots;\theta_N,\epsilon_N)}
    e^{-r\sum_{i=1}^{N} m_{\epsilon_i}\cosh\theta_i}
\end{multline}
where the $\epsilon_i$ labels the particle spectrum of the cBD model.
The rapidity of the $i$-th particle, which is of the type
$\epsilon_{i}$, is $\theta_i$ and has mass $m_{\epsilon_i}$. $F_{Q}$
and $F_{q_{\pm}}$ are the form-factors of exponential fields in an
$N$-particle state with particle spectrum $\epsilon_1,\ldots,
\epsilon_N$. The particle spectrum of the cBD is studied
in~\cite{Smirnov-phi12}. This expression is appropriate to find the
large distance expansion of the correlation functions. The dominant
term is obtained by considering the lightest particle in the spectrum
of the cBD model. For $\Gamma$ small enough (see below) the lightest
particle is the 1-breather. Its mass is
known~\cite{Fateev-Lukyanov-Zamolod2-expect-BD} and it can be
expressed in terms of the Debye length as~\cite{Samaj-asym}
\begin{equation}
  \label{eq:screening-length}
  m=\kappa \left[
    \frac{2\sqrt{3}}{\pi \xi} \sin\left(\frac{\pi\xi}{3}\right)
    \sin\left(\frac{\pi(1+\xi)}{3}\right)
    \right]^{1/2}
\end{equation}
where $\xi=\Gamma/(2-\Gamma)$. The corresponding form-factor
is~\cite{Acerbi-BD-form-fact, Braz-Lukya-form-fact}
\begin{equation}
  \frac{F_Q}{\langle e^{ib Q\phi}\rangle}=
  4\rho \sin\left(\frac{2\pi Q\xi}{3}\right)
  \cos\left(\frac{\pi}{6}\left(1+2\xi-4\xi Q\right)\right)
\end{equation}
with
\begin{equation}
  \rho=i\left[
    \frac{\sin(\pi/3)}{\sin(2\pi\xi/3)\sin(2\pi(1+\xi)/3)}
    \right]^{1/2}
    \exp(I_b/2)
\end{equation}
and
\begin{equation}
I_b= -4 \int_0^{\infty} \frac{\cosh(\frac{t}{6})\sinh(\frac{\xi
t}{3})\sinh(\frac{(1+\xi)t}{3})}{\sinh t \cosh(t/2)} \frac{dt}{t}
\,.
\end{equation}
Notice that the form-factor diverges for $\Gamma=2/3$. This indicates
a change of behavior in the expansion~(\ref{correl-expans}): some
subdominant terms (for $\Gamma<2/3$) become of the same order as the
contribution of the 1-breather at $\Gamma=2/3$
(see~\cite{Tellez-small-r-asympt} for a similar situation in the
short-distance expansion of correlation functions of the sG model). In
the following, we restrict our analysis to $\Gamma<2/3$.

Let us define the effective interaction energy (in units of $k_B T$)
$E_{\lambda,\Gamma q_{\pm}}(r)$ of a microion with charge $2q_{\pm}$
($+2,-1$ respectively) of the electrolyte with the macroion by the
relation $n_{\pm}(r)=n_{\pm}^b \exp[-E_{\lambda,\Gamma
q_{\pm}}(r)]$. At large distances, $r\to\infty$, $n_{\pm}(r)\simeq
n_{\pm}^b(1-E_{\lambda,\Gamma q_{\pm}}(r))$. Using the
expansion~(\ref{correl-expans}) with only the dominant contribution
from the 1-breather form-factor, we obtain, for $mr\gg 1$,
\begin{equation}
  \label{eff-lambda+}
  E_{\lambda,+\Gamma}(r)\sim\frac{8\sqrt{3}}{\pi}\frac{\cos
    \left(\frac{\pi}{6}(1-2\xi)\right)}{\sin\left(\frac{2\pi}{3}(1+\xi)\right)}
  e^{I_b}  
  \sin\left(\frac{2\pi\lambda}{3(2-\Gamma)}\right)
  \cos\left(\frac{2\pi\lambda}{3(2-\Gamma)}-\frac{\pi}{6}
  -\frac{\pi\xi}{3}\right)
  K_0(mr)
\end{equation}
\begin{equation}
  \label{eff-lambda-}
  E_{\lambda,-\Gamma/2}(r)\sim-\frac{8\sqrt{3}}{\pi}
  \frac{\cos\left(\frac{\pi}{6}(1+4\xi)\right)
    \sin\left(\frac{\pi\xi}{3}\right)
  }{\sin\left(\frac{2\pi}{3}(1+\xi)\right)
    \sin\left(\frac{2\pi\xi}{3}\right)}
  e^{I_b}
  \sin\left(\frac{2\pi\lambda}{3(2-\Gamma)}\right)
  \cos\left(\frac{2\pi\lambda}{3(2-\Gamma)}-\frac{\pi}{6}
  -\frac{\pi\xi}{3}\right)
  K_0(mr)
  \,.
\end{equation}
Replacing these expressions into the density profiles and integrating
Poisson equation, we find the effective electrostatic potential $y(r)$
created at a distance $r\gg m^{-1}$ from the macroion. At large
distances, the electrostatic potential takes a similar form to the one
predicted by the linear DH theory $y(r)\sim2\lambda_{\eff}K_0(mr)$,
supporting the hypothesis of the existence of an effective or
renormalized charge $\lambda_{\eff}$. Notice however that there is
also a ``renormalization'' of the screening length: it is $m^{-1}$
given in~(\ref{eq:screening-length}), instead of the usual Debye
length $\kappa^{-1}$. For $0<\Gamma<2/3$, the screening length
$m^{-1}$ is smaller than the Debye length $\kappa^{-1}$. The effective
charge reads
\begin{equation}
  \label{eff-charge}
  \lambda_{\eff}=\frac{\xi\sqrt{3}
    \sin(\pi\xi)e^{I_b}
      \sin\left(\frac{2\pi\lambda}{3(2-\Gamma)}\right)
  \cos\left(\frac{2\pi\lambda}{3(2-\Gamma)}-\frac{\pi}{6}
  -\frac{\pi\xi}{3}\right)
  }{3\sin\left(\frac{2\pi}{3}(1+\xi)\right)
    \sin\left(\frac{2\pi\xi}{3}\right)
    \sin\left(\frac{\pi}{3}(1+\xi)\right)
    \sin\left(\frac{\pi\xi}{3}\right)}
  \,.
\end{equation}
In the low coupling limit $\Gamma\ll 1$ we recover
the results from PB theory~\cite{Trizac-Tellez-Manning,
Tracy-Widom-PB}: $\lambda_{\eff}=\sqrt{3}[2\sin(\frac{2\pi\lambda}{3}
-\frac{\pi}{6})+1]/(2\pi)$ as expected.

%
%
\begin{figure}
  \begin{center}
    \includegraphics[width=\GraphicsWidth]{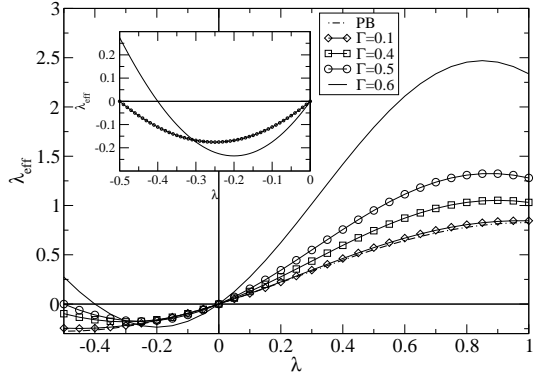}    
  \end{center}
  \caption{
    \label{fig:lambdaeff}
    The effective charge $\lambda_{\eff}$ as a function of the bare
    charge $\lambda$ of the colloid, for different values of the
    coupling constant $\Gamma$ and for the PB approximation. The inset
    details the cases $\Gamma=0.5$ and $\Gamma=0.6$ for
    $\lambda<0$. This later case shows the charge inversion
    phenomenon.  }
\end{figure}
%
%
%
%
%

Figure~\ref{fig:lambdaeff} shows the effective charge as a function of
the bare charge. An important result is that the phenomenon of charge
inversion is possible: when $\Gamma>1/2$ the effective charge and the
bare charge have opposite signs when $-1/2<\lambda<\Gamma-1<0$. The
charge inversion occurs for a highly charged negative macroion which
is overscreened by positive microions of valence $+2$ resulting in a
positive effective charge. On the other hand, for a positive macroion,
screened by microions of valence $-1$ the charge inversion does not
occurs.

The mean field PB formalism does not predict the charge inversion
phenomenon, thus the microions correlations are fundamental for this
phenomenon to take place. The charge inversion has previously been
predicted for a planar charged interface~\cite{Ulander-Kjellander} and
for spherical colloids~\cite{Levesque-Leger} with divalent counterions
and monovalent coions within the hypernetted chain approximation
(HNC). Our model, which is exact (we have solved exactly, without any
approximations, the statistical mechanics of the model), put on more
firm ground the predictions
of~\cite{Ulander-Kjellander,Levesque-Leger} and thus suggests that the
HNC approximation captures adequately the microions correlations
responsible of the charge inversion.

The charge inversion can be actually very large: for example, for
$\Gamma=0.65$ and $\lambda=-0.48$ we have $\lambda_{\eff}=1.22$, thus
a charge inversion ratio of more than 200\% . This is similar to the
``giant'' charge inversion studied in~\cite{Nguyen-giant-charge-inv},
although here the effective charge characterizes the long distance
signature of the potential, contrary to its short-distance behavior as
considered in~\cite{Nguyen-giant-charge-inv}. The charge asymmetry of
the electrolyte seems to be fundamental for the charge inversion to
take place. Exact results for a symmetric electrolyte show no charge
inversion~\cite{Samaj-guest-charges, Tellez-guest-charges}.

The charge inversion is accompanied by a change of behavior in the
large distance behavior of the effective interaction between the
macroion and the microions. For $\lambda>\Gamma-1$, at large
distances, the effective interaction between the macroion and a
counterion is attractive, whereas its interaction with a coion is
repulsive. However, as it can be seen
from~(\ref{eff-lambda+}-\ref{eff-lambda-}), when $\Gamma>1/2$ and
$-1/2<\lambda<\Gamma-1<0$ the interaction of the macroion with a
counterion is now repulsive and its interaction with a coion is
attractive, provided $mr\gg 1$.

A similar change of behavior also occurs at short distances, but only
for the interaction with a coion. We can obtain the short-distance
behavior of the density profiles by using the operator product
expansion (OPE) in the correlation function appearing
in~(\ref{n-correl}). The OPE for the cBD model has been developed
in~\cite{Baseilhac-Stanishkov-descendent-BD}. Using the OPE (see
also~\cite{Hansen-Viot,Tellez-small-r-asympt}) we find, for
$-1/2<\lambda<0$ that $n_{+}(r)\propto r^{4\lambda}$ when $r\to 0$,
yielding an attractive effective interaction
$E_{\lambda,+\Gamma}(r)\sim -4\lambda \ln r$ of the macroion with a
counterion (charge +2) at short distances, which is the expected
behavior. On the other hand the density profile of the coions (charge
$-1$) at short distances exhibits a change of behavior:
\begin{equation}
  \label{n-short}
  n_{-}(r)\propto\
  \begin{cases}
    r^{-2\lambda}&\text{for }\frac{\Gamma-1}{2}<\lambda<0\\
    r^{2(\lambda-\Gamma+1)}&\text{for }
    -\frac{1}{2}<\lambda<\frac{\Gamma-1}{2}<0
  \end{cases}
\end{equation}
giving an effective potential at short distances, $mr\ll1$,
\begin{equation}
  \label{short-dist}
  E_{\lambda,-\Gamma/2}(r)\sim
  \begin{cases}
    2\lambda\ln r&\text{for }\frac{\Gamma-1}{2}<\lambda<0\\
    2(\Gamma-1-\lambda)\ln r&\text{for }
    -\frac{1}{2}<\lambda<\frac{\Gamma-1}{2}<0
    \,.
  \end{cases}
\end{equation}
The change of behavior at $\lambda=(\Gamma-1)/2$ can be understood as
a first step in the counterion
condensation~\cite{Tellez-guest-charges,Tellez-small-r-asympt} for
large coupling $\Gamma$. A fraction of the counterions are condensed
into the macroion so that the effective interaction of the macroion
with the coions at short distances is not anymore the bare Coulomb
potential $2\lambda\ln r$, but it changes to $2(\Gamma-1-\lambda)\ln
r$. The strong correlations between the microions shift the threshold
for counterion condensation from the Manning value $-1/2$ to
$(\Gamma-1)/2$. Since the trademark of the condensation is first
noticed in the change of behavior~(\ref{n-short}) of the coions
density profile, this shift happens only for a situation with added
salt (contrary to the no salt
case~\cite{Naji-Netz-PRL-counterion}). When $\Gamma\to 0$, the above
limit reduces to the standard value $\lambda=-1/2$~\cite{Manning}. In
PB theory, the effective interaction between a coion and the macroion
behave et short-distances as the bare Coulomb potential $2\lambda \ln
r$ down to $\lambda=-1/2$~\cite{Trizac-Tellez-Manning}.

Interestingly, as $\lambda$ decreases beyond $\lambda<(\Gamma-1)/2$
the effective interaction, at short distances, of the macroion with
the coions becomes less and less repulsive. Paradoxically, it even
becomes attractive when $\lambda<\Gamma-1$, provided $\Gamma>1/2$.
The divalent counterions are strongly attracted to the macroion and
since they attract the monovalent coions, there can be a net
attraction of the coions to the macroion. This is accompanied, as we
have seen before, with the charge inversion at large distances. The
very different behavior, at high coupling, of the coion and counterion
density profiles at short distances shows that it is difficult to
define a ``dressed'' charge of the macroion which characterizes the
potential at short distances. Indeed, when $\lambda<(\Gamma-1)/2<0$,
the coions see a dressed charge $\Gamma-1-\lambda$ which can be even
positive, whereas the counterions still see a charge $\lambda<0$.

Returning to the charge inversion phenomenon, we should point out that
it does not necessarily implies attraction between like-charge
macroions. In particular for two identical macroions with charge
$\lambda<\Gamma-1<0$, as both macroions have an inverted effective
charge, the net effect would still be a repulsion between the
macroions.  To study the possibility of like-charge attraction, let us
consider two macroions with linear charge densities $\lambda_1$ and
$\lambda_2$ immersed in the electrolyte at the origin and at $\r$
respectively. In the language of the cBD model, the effective
interaction between the two macroions is given
by~\cite{Samaj-guest-charges,Tellez-guest-charges,Tellez-small-r-asympt}
\begin{equation}
  \exp\left[-E_{\lambda_1\lambda_2}(r)\right]=
  \frac{
    \langle e^{ib Q_1 \phi(0)} e^{ib Q_2 \phi(\r)} \rangle}{
    \langle e^{ib Q_1 \phi}\rangle 
    \langle  e^{ib Q_2 \phi} \rangle}
\end{equation}
with $Q_{1,2}=\lambda_{1,2}/\Gamma$. Using the form-factor theory
explained above we find, for $mr\gg 1$,~\cite{Samaj-asym}
\begin{equation}
  E_{\lambda_1\lambda_2}(r)\sim
  \frac{8\sqrt{3}\,e^{I_b}\, \tilde{\lambda}_1 \tilde{\lambda}_2
    \,K_0(mr)}{\pi
   \sin\left(\frac{2\pi\xi}{3}\right)
   \sin\left(\frac{2\pi(1+\xi)}{3}\right)
  } 
\end{equation}
with
\begin{equation}
  \tilde{\lambda}_{1,2}=
  \sin\left(
  \frac{2\pi\lambda_{1,2}}{3(2-\Gamma)}
  \right)
  \cos\left(\frac{2\pi\lambda_{1,2}}{3(2-\Gamma)}
  -\frac{\pi}{6}-\frac{\pi\xi}{3}\right)\,.
\end{equation}
The ``charges'' $\tilde{\lambda}_{1,2}$ exhibit the same changes of
sign as the effective charge~(\ref{eff-charge}) shown in
Fig.~\ref{fig:lambdaeff}. Thus, two negatively charged macroions can
have an attractive effective interaction at large distances provided
that $\lambda_1<\Gamma-1<0$ and $\Gamma-1<\lambda_2<0$ for
$\Gamma>1/2$. Also a positive macroion (say $\lambda_2>0$) can have a
repulsive effective interaction with a negative macroion (charge
$\lambda_1$) provided $\lambda_1<\Gamma-1<0$.


Summarizing, we have considered a model that describes the effective
charge of cylindrical macroions immersed in a charge-asymmetric
$+2$/$-1$ electrolyte solution. This model is exactly solvable and
predicts charge inversion at high coupling $\Gamma>1/2$ for negatively
charged macroions provided their linear charge density
$\lambda<\Gamma-1<0$. Also, we have shown that a shift in the counterion
condensation threshold from $\lambda=-1/2$ to $\lambda=(\Gamma-1)/2$
occurs due to the strong coupling between the microions, which is part
of the mechanism of charge inversion.


\acknowledgments Financial support from ECOS-Nord/COLCIENCIAS,
COLCIENCIAS (1204-05-13625) and Comit\'e de Investigaciones, Facultad
de Ciencias, Universidad de los Andes is acknowledged. The author
thanks L.~\v{S}amaj for interesting discussions on the model, and
M.~Camargo for a useful remark.


\end{document}